\documentclass{article}
\usepackage{amsfonts,amssymb}
\usepackage[noblocks]{authblk}
\usepackage{amsthm}
\usepackage{geometry,algorithm}  
\usepackage{algorithmic}
\usepackage{amsmath} 
\usepackage{cite}  
\usepackage{graphicx}  
\usepackage{caption}
\usepackage{subfigure}
\usepackage{subfigure}
\usepackage{subfigure}
\usepackage{float}
\usepackage{booktabs}
\usepackage{multirow}
\usepackage{threeparttable}
\usepackage{caption}
\usepackage[utf8]{inputenc}
\usepackage[T1]{fontenc}
\usepackage{booktabs}
\usepackage{array, caption, threeparttable}
\usepackage[font=small,labelfont=bf,labelsep=none]{caption}
\usepackage{tikz}
\newcommand*{\circled}[1]{\lower.7ex\hbox{\tikz\draw (0pt, 0pt)%
		circle (.5em) node {\makebox[1em][c]{\small #1}};}}
\begin{document}
	\renewcommand{\thefootnote}{\fnsymbol{footnote}}
	\captionsetup[figure]{labelfont={bf},name={Fig },labelsep=period}
	\title{\textbf {Certification  of three black boxes with unsharp measurements using $3 \rightarrow 1 $ sequential quantum random access codes} }	
	
\author[1,2]{Shihui Wei}
\author[1,2]{Fenzhuo Guo}
\author[1]{Fei Gao}
\author[1]{Qiao-Yan Wen}
\affil[1]{
	\emph{State Key Laboratory of Networking and Switching Technology, Beijing University of  Posts and Telecommunications, Beijing 100876, China}}
\affil[2]{
	\emph{School of Science, Beijing University of Posts and Telecommunications, Beijing 100876, China}}	
\maketitle

{\noindent\textbf {Abstract}\quad}{	Unsharp  measurements play an increasingly important role in quantum information theory. In this paper, we study a three-party  prepare-transform-measure experiment with unsharp  measurements  based on $ 3 \rightarrow 1 $ sequential  random access codes (RACs). We derive optimal trade-off  between the two correlation witnesses in $ 3 \rightarrow 1 $ sequential quantum random access codes (QRACs), and use the result  to complete  the self-testing of quantum  preparations, instruments and measurements  for  three sequential  parties. We also give the upper and lower bounds of the sharpness parameter to complete the robustness analysis of the self-testing scheme.
	In  addition, we find  that  classical correlation witness violation based on $3 \rightarrow 1 $ sequential RACs  cannot be obtained  by both  correlation witnesses simultaneously. This means that if the second party uses strong unsharp measurements  to overcome the classical upper bound, the third party cannot do so even with sharp measurements.
	Finally, we   give the  analysis and comparison  of the  random number generation efficiency under different sharpness parameters based on the  determinant value,  $2 \rightarrow 1 $ and  $3 \rightarrow 1 $ QRACs  separately. This letter sheds new light on generating  random numbers among multi-party in semi-device independent  framework.}

	\section{Introduction}
	It is an important task of quantum mechanics to obtain relevant information  of the system by quantum measurement \cite{heisenberg1949physical}. Sharp measurements, obtaining the maximum amount of information about the system, collapse the quantum states   randomly into one  of the eigenstates of the measured observable.
	On the other hand, unsharp  measurements  cause little disturbance to the system and allow us to obtain partial   information of the system.  Unsharp measurements that aim at bringing minimal disturbance to the system are often called “weak measurements” \cite{aharonov1988result}.
	Unsharp, and especially weak measurements play an important role in many quantum information processing tasks, for instance, quantum random number generation \cite{curchod2017unbounded,li2018three,an2018experimental,coyle2018one}, state tomography \cite{lundeen2012procedure,wu2013state}, sequential quantum correlations \cite{silva2015multiple,shenoy2019unbounded,anwer2019noise,brown2020arbitrarily} and others.
	
	Recently, Mohan  $ et$  $al$.\cite{mohan2019sequential}  discussed unsharp measurements in a three-party prepare-transform-measure experiment with  $2 \rightarrow 1 $   random access codes (RACs) \cite{ambainis1999dense,ambainis2008quantum}, which has  been  experimentally demonstrated \cite{anwer2020experimental, foletto2020experimental}.
	In Ref.\cite{mohan2019sequential},
	they characterized optimal trade-off  between the two $ 2 \rightarrow 1 $ quantum random access codes (QRACs),  and applied the result to realise semi-device independent (SDI) self-testing of quantum measurement instruments. 
	Self-testing  is the task of characterizing unknown quantum states and measurements solely from the measurement statistics.
	
	The original idea of testing states and measurements was proposed by Popescu  $ et$  $al.$\cite{popescu1992states} based on  Clauser-Horne-Shimony-Holt (CHSH) inequality \cite{clauser1969proposed}. If we obtain
	the maximal violation of the CHSH inequality, we can determine that the state  measured is a two-qubit maximally entangled one and the measurements are two anticommuting Pauli observables. Since then, a growing number of  quantum states have also been proved to be self-testable \cite{yang2013robust,wu2014robust,pal2014device,kaniewski2017self,coladangelo2017all,baccari2020scalable}. All of the above self-testing  schemes  rely on quantum nonlocality within the device independent framework.  Beyond those based on nonlocality,  Tavakoli  $ et$  $al$. presented a self-testing method for quantum prepare-and-measure experiment in 2018 \cite{tavakoli2018self}. After that, various self-testing schemes for different quantum states have been proposed under such framework  \cite{mohan2019sequential,farkas2019self,mironowicz2019experimentally,wei2019robustness,miklin2020semi,tavakoli2020self,miklin2020universal,tavakoli2020semi}. These  schemes consider quantum systems in fixed dimensions and belong to SDI framework,  which opens interesting possibilities for quantum information processing.
	
	In this paper, we study a three-party  prepare-transform-measure experiment with $ 3 \rightarrow 1 $ sequential RACs, in which three black boxes in the safe area have  been considered. For convenience, we call these  black boxes Alice, Bob and Charlie sequentially.
	Alice randomly  prepares one of the quantum states and sends it to Bob. Bob applies quantum instrument on it, and gets both a classical and quantum output, then Bob sends his post-measurement state to Charlie who will make  further measurement. In our scenario,   Alice encodes a three-bit long random sequence into an  one-bit message  while both Bob and Charlie aim to decode any of the three bits held by Alice, i.e. they individually implement a $  3 \rightarrow 1 $ RAC with Alice. 
	To  analyze the optimization problem on the third party's correlation witness,	
	we derive optimal trade-off   between the two correlation witnesses in $ 3 \rightarrow 1 $ sequential QRACs.
	This result allows us to self-test  Alice’s preparations, Bob’s instruments and  Charlie’s measurements within the SDI framework.
	Moreover,  if the  pair of these two correlation witnesses is suboptimal (here,  we can regard it as the deviation caused by noise which is characterized as  the sharpness parameter of Bob's quantum measurement instruments),
	we give the upper and lower bounds of the sharpness parameter, and complete the robustness analysis of the self-testing scheme.
	
	Whether the three-party protocol can be extended to any number of parties has always been a hot research topic in recent years  if  all transformers use weak measurements \cite{mal2016sharing,maity2020detection}.
	We find that  the double  classical correlation witness violation cannot be obtained   based on $3 \rightarrow 1 $ sequential QRACs. If Bob  uses unsharp measurements strong enough to overcome the classical upper bound, Charlie cannot do so even with maximal strength.
	Finally, the analysis of all our results sheds new light on the interaction between the three-party quantum dimension witness and the unsharp  measurement technology. We apply it to the generation of SDI quantum random numbers,  and give the local randomness on Bob's side and Charlie's side respectively.
	We   give the  analysis and comparison  of the  random number generation efficiency under different sharpness parameters based on the  determinant value,  $2 \rightarrow 1 $ and  $3 \rightarrow 1 $ QRACs  separately.
	
	\section{The $ 3 \rightarrow 1 $ sequential RACs }
	To better explain our theory, we first introduce a three-party prepare-transform-measure experiment based on $  3 \rightarrow 1 $ sequential RACs in detail.
	
\begin{figure}
		\centering
		\includegraphics[width=4in]{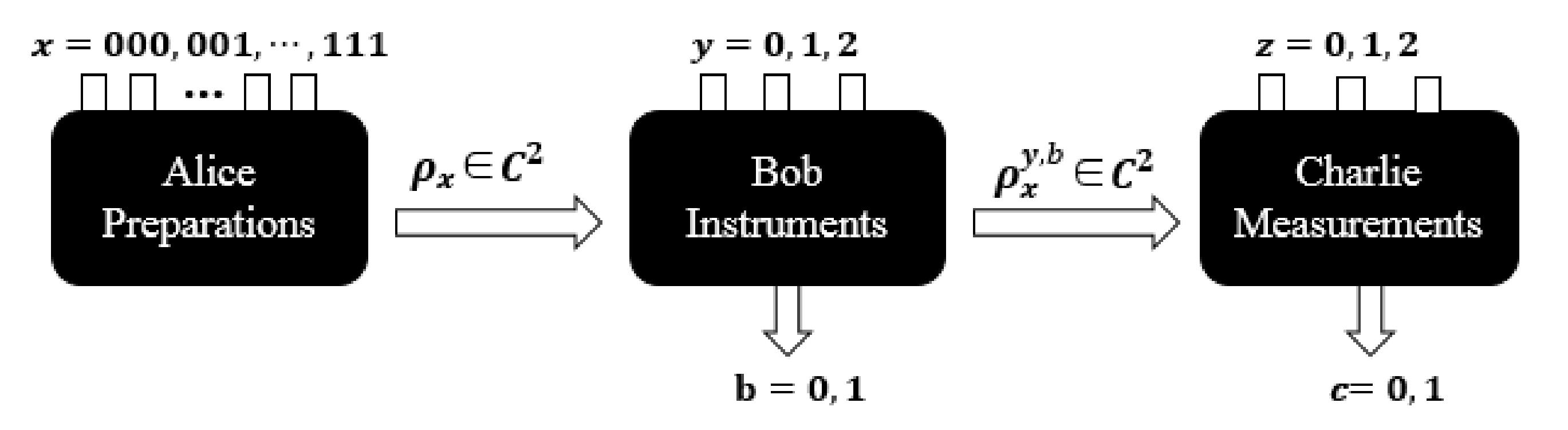}
		\caption{The three-party prepare-transform-measure experiment with  $  3 \rightarrow 1 $ sequential RACs. Alice  prepares a qubit $ \rho_{x} $ according to her three classical  bits $x$ ($ x\in\{000,001,\cdots,111\} $)  and sends it to Bob.  Bob performs his measurement instrument on the received qubit depending on his input $y$  $( y\in\{0,1,2\}) $ and gets the measurement result denoted as $b$ ($ b\in\{0,1\}$) and a qubit post-measurement state $ \rho^{y,b}_{x}$. Charlie performs his measurement on $ \rho^{y,b}_{x}$ depending on his input $z$  ($z\in\{0,1,2\} $) and gets the measurement result denoted as $c$ ($ c\in\{0,1\}$).}
		\label{fig:1}       
	\end{figure}

	In our experiment,
	Alice has the freedom to choose one of eight preparations $\{\rho_{x}\} $ where $x=x_0x_1x_2$ ($ x_{0},x_{1},x_{2}\in\{0,1\} $), but knows nothing about these quantum states apart from their dimensionality 2.
	For a given input $ x $, Alice prepares a quantum state $ \rho_{x}$ and  sends it to Bob.  Bob performs one of three quantum measurement instruments on $ \rho_{x}$ based  on his input $ y  $ ($y\in\{0,1,2\}$), and  gets both a classical binary outcome $b$ $ (b\in\{0,1\}) $  and a qubit output $ \rho^{y,b}_{x}$ [11]. Since  these quantum measurement instruments are completely positive trace-preserving maps, we  characterise the quantum measurement instrument by Kraus operators $ \{K_{b\mid y}\}$. Therefore, we obtain the Bob's qubit post-measurement state
	\begin{eqnarray}
	\rho^{y,b}_{x}=\frac{K_{b\mid y}\rho_{x}K^{\dagger}_{b\mid y}}{\mathrm{tr} (K^{\dagger}_{b\mid y}K_{b\mid y}\rho_{x})}.
	\end{eqnarray}
	Then Bob sends the qubit post-measurement state $ \rho^{y,b}_{x}$ to Charlie, Charlie performs one of three   sharp  measurements   $\{ C_{z}\}$  on $ \rho^{y,b}_{x}$ depending on his input $z$ ($z\in\{0,1,2\} $) and gets a measurement result denoted as $ c$ ($ c\in\{0,1\}$). All the random bits $ x_{0},x_{1},x_{2}, y $ and $z$ are independent and uniformly distributed.
	This scenario is schematically depicted in  Fig.\ref{fig:1}.
	
	After repeating this procedure many times, Alice, Bob and Charlie can estimate the conditional probability distribution $p(b,c| x,y,z) $ =tr$[K_{b\mid y}\rho_{x}K^{\dagger}_{b\mid y}C_{c| z}]$, which denotes the probability  of Bob and Charlie obtaining the outcome $ b,c $ when the Kraus operators $ \{K_{b\mid y}\}$ and  measurements  $ \{C_{c\mid z}\}$  are  performed on Alice's prepared state $ \rho_{x}$ sequentially.
	
	Alice encodes three bits of information  into a single bit of information  while both Bob and Charlie aim to recover some randomly chosen subset of the data held by Alice, i.e. they individually implement a $  3 \rightarrow 1 $ RAC with Alice.
	We are interested in two separate correlation witnesses in $ 3 \rightarrow 1 $ sequential RACs. The correlation witness is the average guessing probability. The two respective average success rates read
	\begin{eqnarray}
	\mathcal{A}_{AB}= \frac{1}{24} \sum_{x_0,x_1,x_2,y}p(b=x_{y}|x_0x_1x_2,y).
	\end{eqnarray}
	\begin{eqnarray}
	\mathcal{A}_{AC}= \frac{1}{24} \sum_{x_0,x_1,x_2,z}p(c=x_{z}|x_0x_1x_2,z).
	\end{eqnarray}
	This means that, upon receiving input $ y$ ($ z$), Bob’s (Charlie’s)   measurement
	device should get the output  $ b=x_{y}  $ ($ c=x_{z} $), i.e., the $y$-th ($ z$-th)  of the input bit-string $x$ received by the Alice’s preparation device.
	
	Let's firstly consider the  $  3 \rightarrow 1 $ RAC between Alice and Bob. Since Bob’s  quantum instrument (a completely positive trace-preserving map) realises a measurement,  the Kraus operators  must satisfy the completeness relation $\forall y:B_{0\mid y}+B_{1\mid y}=I, $ where $ B_{b\mid y}=K^{\dagger}_{b\mid y}K_{b\mid y}$ are the corresponding elements of the positive operator-valued measures (POVMs). Moreover,  the observable  $ B_{y}$ is defined as $ B_{y}=B_{0| y}-B_{1| y} $. Therefore, we have
	\begin{eqnarray}
	\mathcal{A}_{AB}=\frac{1}{24} \sum_{x,y}\mathrm{tr} [K_{x_{y}| y}\rho_{x}K^{\dagger}_{x_{y}| y}]
	=\frac{1}{24} \sum_{x,y}\mathrm{tr} [\rho_{x}B_{x_{y}| y}].\label{eq4}
	\end{eqnarray}
	
	In a two-dimensional system, the upper bound of $ \mathcal{A}_{AB} $ corresponding to the quantum system  is   $ \mathcal{A}_{AB}^{Q}=\frac{1}{2}+\frac{1}{2\sqrt{3}}\approx0.79$ \cite{navascues2015bounding}, and the maximum value of $ \mathcal{A}_{AB} $ corresponding to the classical system is $\mathcal{A}_{AB}^{C}=\frac{3}{4}. $ The quantum bound  $ \mathcal{A}_{AB}^{Q}$ can be  obtained via the following  “$ ideal $” strategy.    Alice’s eight  preparations  are chosen as
	\begin{eqnarray}
	\rho_{x_0x_1x_2}^{ideal}=\frac{I+\vec{n}_{x_0x_1x_2}^{ideal}\cdot\vec{\sigma}}{2},\label{eq5}
	\end{eqnarray}
	where $ \vec{n}_{x_0x_1x_2}^{ideal}=\frac{1}{\sqrt{3}}((-1)^{x_{0}},(-1)^{x_{1}},(-1)^{x_{2}})$ is Bloch vector, $\vec{\sigma}=(\sigma_{x},\sigma_{y},\sigma_{z})$ denotes the Pauli matrix vector. 
	To achieve the maximum value
	$ \frac{1}{2}+\frac{1}{2\sqrt{3}}$, the corresponding measurements in Bob’s side are:
	
	\begin{eqnarray}
	B_{0}^{ideal}=\sigma_{x},  B_{1}^{ideal}=\sigma_{y},   B_{2}^{ideal}=\sigma_{z}.
	\label{eq6}
	\end{eqnarray}
	
	This means that Alice’s preparations  are pure states, and correspond to the eigenvector of  $L_{x}=\sum_{x}(-1)^{x_{y}}B_{y}^{ideal}    $ associated to its largest eigenvalue. Such a set of preparations correspond  to Bloch vectors forming a cube
	on the Bloch sphere and  the measurements correspond to three mutually unbiased bases (i.e., three pairwise anticommuting Pauli observables).
	It should be emphasized that this set of quantum states and measurements is   uniquely determined in the sense of a unitary and a complex conjugation.
	

	Correspondingly, the  $ 3 \rightarrow 1 $ RAC  between Alice and Charlie is considered.
	Charlie  can  also guess the average probability of success of a bit of Alice.
	In a classical model, the state at all times is diagonal in the same basis. Bob can interact with  Alice’s preparations  without disturbing her states. In this case, $ \mathcal{A}_{AB},\mathcal{A}_{AC}\in[\frac{1}{2},\frac{3}{4}]$.
	
	However, in a quantum model, Bob’s instrument disturbs the physical state of Alice’s qubit, and
	therefore he cannot relay Alice’s original quantum message to Charlie. The effective state $ \tilde{\rho}_{x} $ received by Charlie is the post-measurement state of Bob averaged
	over $y$ and $b$, we write
	\begin{eqnarray}
	\tilde{\rho}_{x}=\frac{1}{3}\sum_{y,b}p(b| y) \rho_{x}^{y,b}=\frac{1}{3}\sum_{y,b}K_{b\mid y}\rho_{x}K^{\dagger}_{b\mid y}.
	\end{eqnarray}
	Therefore, we have
	\begin{eqnarray}
	\mathcal{A}_{AC}=\frac{1}{24} \sum_{x,z}\mathrm{tr} [\tilde{\rho}_{x}C_{x_{z}| z}]= \frac{1}{72} \sum_{x,y,b,z}\mathrm{tr} [K_{b\mid y}\rho_{x}K^{\dagger}_{b\mid y}C_{x_{z}| z}].\label{eq8}
	\end{eqnarray}
	
	In this case, $ \mathcal{A}_{AB},\mathcal{A}_{AC}\in[1/2,(1+1/\sqrt{3})/2]$. Evidently, $\mathcal{A}_{AB}$ is independent of Charlie. However, $\mathcal{A}_{AC}$ is not independent of Bob because he operates on Alice’s original preparation that reaches Charlie. In other words, Charlie’s ability to access the desired information depends on Bob’s preceding interaction. 
	Furthermore, we are interested in the relation between  $\mathcal{A}_{AB}$ and  $\mathcal{A}_{AC}$. We intuitively know that in order to make the value of  $\mathcal{A}_{AB}$ larger, the quantum states prepared by Alice should be close to the quantum states in equation (\ref{eq5}), and Bob's quantum measurement instruments should be close to the measurements in equation (\ref{eq6}) in the sense of  local isometry.
	This means that Bob’s measurements should be reasonably sharp. Once $ \mathcal{A}_{AB}$  achieves the maximum value  $ \frac{1}{2}+\frac{1}{2\sqrt{3}}$,
	we can know that the disturbance caused by Bob to the original quantum state prepared by Alice is maximal, and accordingly, the value of $ \mathcal{A}_{AC}$ that we can obtain should be minimal. The converse is also true. It is therefore natural to ask  what is the  optimal trade-off  between  the two correlation witnesses in a three-party prepare-transform-measure experiment. We derive optimal trade-off relation between the two correlation witnesses based on $ 3 \rightarrow 1 $ sequential QRACs in the next section.
	\section{Optimal trade-off  between the two correlation witnesses based on $ 3 \rightarrow 1 $ sequential QRACs}
	In this section,  we analyze  what values are attainable for the pair of the two correlation witnesses  ($\mathcal{A}_{AB}$,  $\mathcal{A}_{AC}$) based on $ 3 \rightarrow 1 $ sequential QRACs in detail. From the above discussion,  we can rephrase  the problem as follows: for a given value  of $\mathcal{A}_{AB}\in [\frac{1}{2}, \frac{1}{2}(1+\frac{1}{\sqrt{3}})]$, what is the optimal value of $\mathcal{A}_{AC}$ in quantum theory?
	We will solve this problem by considering the related optimization problem
	\begin{eqnarray}
	\mathcal{A}_{AC}^{\mathcal{A}_{AB}}=\max_{\rho,U,M,C}\mathcal{A}_{AC}  \cr
	\mbox{ subject    to}: \cr
	\forall x:\rho_{x}\in C^{2}, \quad \rho_{x}\geq 0,\quad \mathrm{tr}\rho_{x}=1, \cr
	\forall y,b:U_{yb}\in \mbox { SU(2)},\quad B_{b\mid y}\geq 0,\quad  B_{0\mid y}+B_{1\mid y}=I,\cr
	\forall z,c:C_{c\mid z}\geq 0,\quad  C_{0\mid z}+C_{1\mid z}=I, \cr
	\frac{1}{2}\leq\mathcal{A}_{AB}\leq\frac{1}{2}(1+\frac{1}{\sqrt{3}}).
	\end{eqnarray}
	The optimization of the process takes over all Alice’s  preparations, Bob’s instruments and Charlie’s measurements. We solve this optimization problem by first giving a lower bound on $\mathcal{A}_{AC}^{\mathcal{A}_{AB}}$ and then matching it with an upper bound.
	We can obtain the optimal value $\mathcal{A}_{AC}^{\mathcal{A}_{AB}}$ and give the following proposition.
	
	$\mathbf{Proposition}$ $\mathbf{1}$. The optimal trade-off between the pair  of the two correlation witnesses  ($\mathcal{A}_{AB}$,  $\mathcal{A}_{AC}$) based on $ 3 \rightarrow 1 $ sequential QRACs corresponds to
	\begin{eqnarray}
	\mathcal{A}_{AC}^{\mathcal{A}_{AB}}=\frac{1}{2}+\frac{\sqrt{3}}{18}(1+2\sqrt{12\mathcal{A}_{AB}-12\mathcal{A}_{AB}^{2}-2}),
	\label{eq10}
	\end{eqnarray}
	where $ \mathcal{A}_{AB}\in[1/2,(1+1/\sqrt{3})/2] $.

	\begin{figure}
		\centering
		\includegraphics[width=3in]{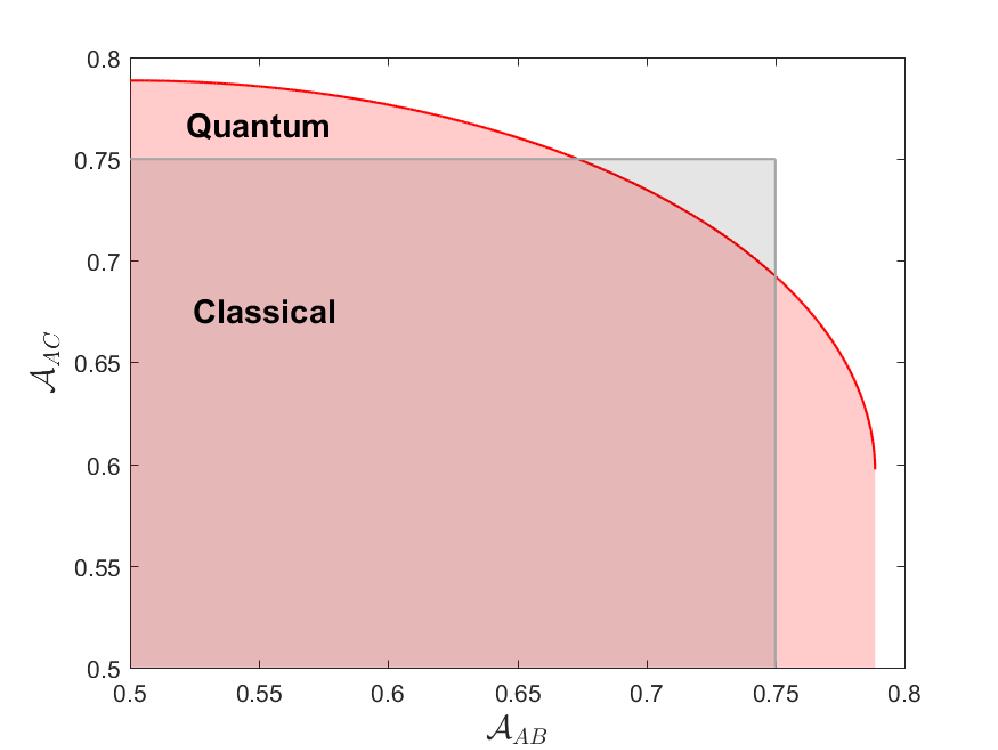}
		\caption{The trade-off relations between the two $ 3 \rightarrow 1 $ RACs in a classical and quantum model respectively.}
		\label{fig2:2}       
	\end{figure}
	
\begin{figure}
		\centering
		\includegraphics[width=3in]{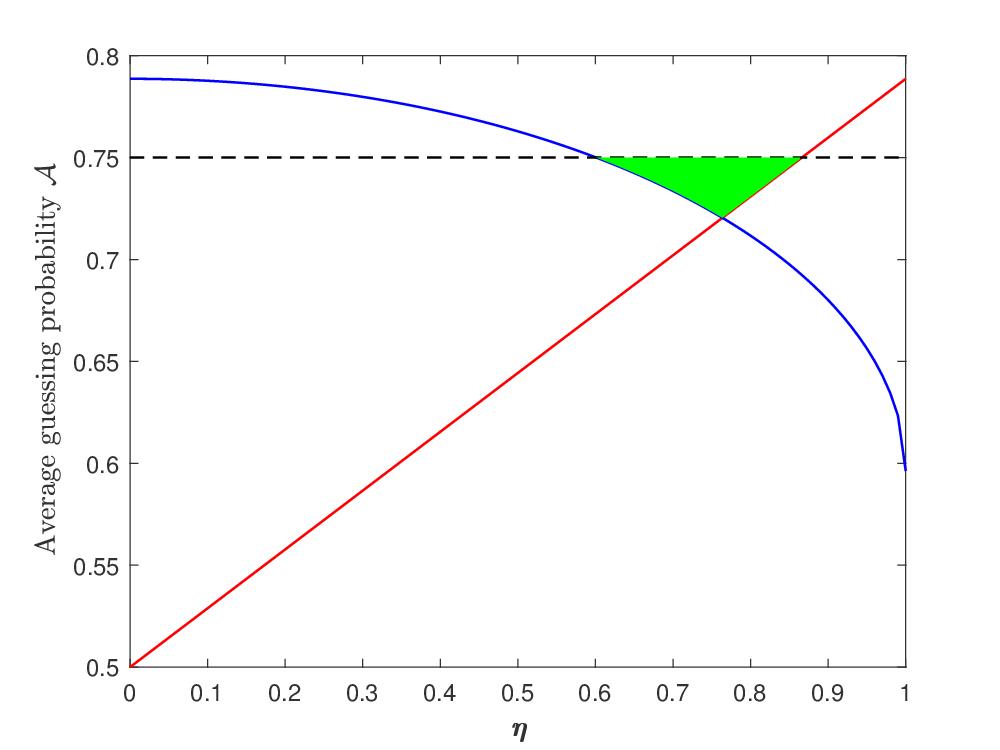}
		\caption{Average guessing probability (Correlation witness) $ \mathcal{A}$  as a function
			of the sharpness parameter $ \eta $.
			The red line is the correlation witness	between Alice and Bob. The blue line is  the correlation witness	between Alice and Charlie. The green area represents that  neither  $ \mathcal{A}_{AB}$  nor $ \mathcal{A}_{AC}$  can realize  the classical correlation witness violation.}
		\label{fig:3}       
	\end{figure}

	The proof of Proposition 1	is too lengthy to be included in the main text and is given in	Appendix A.
	The respective  trade-offs  between the two correlation witnesses based on $ 3 \rightarrow 1 $  RACs in a classical and quantum model  are given in Fig.\ref{fig2:2}. We should point out that all the presented bounds are tight in the sense that there exist states and measurements reaching these average success probabilities. In particular, when $ \mathcal{A}_{AB}$ reaches the  maximum  value $ \frac{1}{2}+\frac{1}{2\sqrt{3}}$, the value of $ \mathcal{A}_{AC}$ is $ \frac{1}{2}(1+\frac{\sqrt{3}}{9})>\frac{1}{2}$,  that is, Charlie’s average guessing probability is still better than random guessing. In this case, the effective ensemble relayed by Bob (the first decoder) to Charlie (the second decoder) corresponds to that originally prepared states (given in equation (\ref{eq5})) by Alice, but with Bloch vectors of $\frac{1}{3}$ the original length. Moreover, if Charlie performs the same Kraus operators (given in equation (\ref{eq6})) as Bob, we find $\mathcal{A}_{3}=\frac{1}{2}(1+\frac{\sqrt{3}}{27})$. Similarly, the effective ensemble  to David (the third decoder) relayed by  Charlie  will be identical to that relayed by Alice, except that the Bloch vectors will be shrunk to $ \frac{1}{9}$ of the original length. Continuing the sequence in this manner, we find that the worst average guessing probability obtained by the $ k$-th decoder is
	\begin{eqnarray}
	\mathcal{A}_{k}=\frac{1}{2}(1+\frac{\sqrt{3}}{3^{k}}).
	\end{eqnarray}
	In addition, we find that both $ 3 \rightarrow 1 $ QRACs cannot always outperform the    $ 3 \rightarrow 1 $ classical RACs.  In order to understand  these two pairs of correlation witnesses ($\mathcal{A}_{AB}$,  $\mathcal{A}_{AC}$) under $ 3 \rightarrow 1 $ classical RACs and QRACs more intuitively,  the average success probability as a function of the sharpness parameter $ \eta $ is
	studied and shown in  Fig.\ref{fig:3}. Note that the green area in  Fig.\ref{fig:3} corresponds to the gray area above the red line in  Fig.\ref{fig2:2}, which represents that  neither  $ \mathcal{A}_{AB}$  nor $ \mathcal{A}_{AC}$  can realize  the classical correlation witness  violation. Furthermore, $ \mathcal{A}_{AB}$ and  $ \mathcal{A}_{AC}$ cannot  achieve the double classical correlation witness violation simultaneously. This implies that if Bob  uses unsharp measurements strong enough to   achieve the maximal classical  violation of the correlation witness, Charlie cannot do so even with maximal strength. This is entirely different from the situation of $ 2 \rightarrow 1 $ sequential QRACs \cite{mohan2019sequential}.

	To understand why this is so, we need to consider the constructions of the optimal $2 \rightarrow 1 $ and $3 \rightarrow 1 $ QRACs in Ref.\cite{ambainis2008quantum}. In a two-dimensional system, both the states and the measurements can be represented by the unit vectors on Bloch sphere. From Ref.\cite{ambainis2008quantum}, we notice that compared with the optimal $3 \rightarrow 1 $ QRAC, the states and the measurements for the optimal $2 \rightarrow 1 $ QRAC lie in one plane and do not use the full size of the  space.  
	Our result seems to be match with the results in Ref.\cite{li2012semi}. Li $ et$  $al$. calculated the dependence of the effectiveness of the randomness generation on $n$ ($n \rightarrow 1 $ QRAC) and found it optimal for $n = 3$, and provided a similar explanation for this fact. 
	Here,  we consider a case where the most information can be obtained with the least interference with the original quantum state of the system.  In order to understand our result intuitively, we can consider the following scenario: the maximum amount of randomness generated (the maximum amount of information) is obtained when $n=3$, Bob causes the most disturbance to the original quantum state, resulting in the minimum amount of information obtained by Charlie's measurement (i.e., $ \mathcal{A}_{AC} $ is the minimum). Besides, the upper bound of $ \mathcal{A}_{AB}^{2 \rightarrow 1} (\mathcal{A}_{AB}^{3 \rightarrow 1}) $ corresponding to the quantum system based on $2 \rightarrow 1 (3 \rightarrow 1) $ QRAC   is approximately $ 0.85   (0.79)$. The classical upper bound of both is the same, which is 0.75. Obviously, compared with $2 \rightarrow 1 $ RAC, the interval between quantum upper bound and classical upper bound is narrower under $3 \rightarrow 1 $ RAC. Therefore,  compared with the  situation of $ 2 \rightarrow 1 $ sequential RACs, it seems reasonable  that  classical correlation witness violation based on $3 \rightarrow 1 $ sequential RACs  cannot be obtained  by both  correlation witnesses simultaneously.
	
	This is expected to spark widespread interest in more general scenarios in future studies. For example, whether this property exists for  $n \rightarrow 1 $   QRACs and more general scenarios (higher-dimensional and many-input QRACs, as well as longer sequences of observers). It is potentially useful in recycling quantum resources in the context of various information processing tasks. Besides, this work also has played an important role in promoting the study of sequential QRACs and has far-reaching influence in related fields.  
	\section{Self-testing and  robustness analysis of three black boxes with  $ 3 \rightarrow 1 $ sequential QRACs}
	According to the derivation of the optimal trade-off relation between the two correlation witnesses based on $ 3 \rightarrow 1 $ sequential QRACs in the previous sections, we find that this process implies a set of self-testing schemes for Alice’s preparations, Bob’s instruments and  Charlie’s measurements  under the SDI framework.
	In order to obtain a self-testing
	scheme, we must additionally prove that the optimal correlation witness pair $(\mathcal{A}_{AB}, \mathcal{A}_{AC}^{\mathcal{A}_{AB}})$ allows only one implementation with a specific  set of Alice’s preparations, Bob’s instruments and  Charlie’s measurements  (up to the collective unitary transformation). We will discuss it in detail below.

	Firstly, we have already shown that Alice’s preparations should be pure and pairwise antipodal. In Lemma 1, we obtain $ \mu= \arccos \frac{1}{\sqrt{3}},\varphi=\frac{\pi}{4}$. Thus, Alice’s preparations correspond to a cube  on the surface of the Bloch sphere. The above arguments fully characterise Alice’s preparations up to a reference frame.
	Next, Bob’s instrument  realises a measurement $ B_{y}=\alpha_{y}I+\vec{t}_{y}\cdot\vec{\sigma} $ where $\vec{t}_{y}=(t_{y0},t_{y1},t_{y2}). $ From Lemma 1, we get $ \alpha_{y}= 0$ and
	$\label{cases}
	t_{yj}=\begin{cases}\eta&y=j\\
		0&y\neq j\end{cases}$, where $ y,j\in \{0,1,2\}$. Therefore,
	$ B_{0}=\eta\sigma_{x}, B_{1}=\eta\sigma_{y},B_{2}=\eta\sigma_{z} $. we also get $\gamma_{0}=\sigma_{x},\gamma_{1}=\sigma_{y},\gamma_{2}=\sigma_{z}$. Moreover, we have made the optimal choice of letting $ V_{yzb}=U_{yb}^{\dagger}C_{0| z}U_{yb} $ project onto the eigenvector of $\sqrt{B_{b| y}}\gamma_{z}\sqrt{B_{b| y}} $ with the largest eigenvalue $ \lambda_{\max} $. Thus,
	$ V_{y0b}=U_{yb}^{\dagger}C_{0| 0}U_{yb}=| +  \rangle\langle + | $,
	$ V_{y1b}=U_{yb}^{\dagger}C_{0| 1}U_{yb}=| \imath  \rangle\langle \imath | $,
	$ V_{y2b}=U_{yb}^{\dagger}C_{0| 2}U_{yb}=| 0  \rangle\langle 0 | $,
	where $ | +  \rangle=\frac{1}{\sqrt{2}} (| 0  \rangle + | 1  \rangle), | \imath  \rangle=\frac{1}{\sqrt{2}} (| 0  \rangle + i| 1  \rangle).   $
	Hence, For Charlie’s measurements $  C_{z}=C_{0\mid z}-C_{1\mid z}$, we get
	$ \forall y,b:U_{yb}=U $    and
	$ C_{0}=U\sigma_{x}U^{\dagger}, C_{1}=U\sigma_{y}U^{\dagger}, C_{2}=U\sigma_{z}U^{\dagger} $.
	Finally, We are ready to present the result, which is given by the following  proposition.

	$\mathbf{Proposition}$ $\mathbf{2}$.  Once  the optimal trade-off between the pair  of the two correlation witnesses  ($\mathcal{A}_{AB}$,  $\mathcal{A}_{AC}$) based on $ 3 \rightarrow 1 $ sequential QRACs is obtained, we can implement the self-test of the following  unique Alice’s preparations,   Bob’s instruments and  Charlie’s measurements (up to collective unitary transformations)
	
	(a)  Alice’s states are pure, which correspond to a set of the states forming a cube on the surface of the Bloch sphere. These eight states are  given in equation (\ref{eq5}).
	
	(b) Bob’s instruments are Kraus operators $K_{b\mid y} = U_{yb}\sqrt{B_{b| y}} $, where  $ B_{b\mid y}=K^{\dagger}_{b\mid y}K_{b\mid y}$ are the corresponding elements of the positive operator-valued measures (POVMs), $B_{y}=B_{0\mid y}-B_{1\mid y}=K^{\dagger}_{0\mid y}K_{0\mid y}-K^{\dagger}_{1\mid y}K_{1\mid y}. $ Specifically, $ \forall y,b:U_{yb}=U, B_{0}=\eta\sigma_{x}, B_{1}=\eta\sigma_{y},B_{2}=\eta\sigma_{z} $ where $ \eta=\sqrt{3}(2\mathcal{A}_{AB}-1)$.

	(c) Charlie’s measurements are rank-one sharp measurements, where
	$ C_{0}=U\sigma_{x}U^{\dagger}, C_{1}=U\sigma_{y}U^{\dagger}, C_{2}=U\sigma_{z}U^{\dagger} $.

	This set of strategies corresponds to the red line in Fig.\ref{fig2:2}.
	The unitarity of these operators in Proposition 2  holds only if the statistics are ideal, however, we can never have perfect statistics in the real case. An interesting question is how to make this  result  have noise-tolerance. To solve this problem, we can bound the sharpness parameter  of Bob’s instruments from noisy correlations. We  rewrite the equation (\ref{eq21}) as
	
	\begin{eqnarray}
	\mathcal{A}_{AB}=\frac{1}{2}+\frac{1}{24}( | \vec{s}_{0}|  | \vec{t}_{0}|\vec{\hat{s}}_{0}\cdot\vec{\hat{t}}_{0} +  | \vec{s}_{1}|  | \vec{t}_{1}|\vec{\hat{s}}_{1}\cdot\vec{\hat{t}}_{1} + | \vec{s}_{2}|  | \vec{t}_{2}|\vec{\hat{s}}_{2}\cdot\vec{\hat{t}}_{2}),
	\end{eqnarray}
	where $\vec{\hat{s}},$ $\vec{\hat{t}} $ are the normalized form of $ \vec{s},$ $\vec{t} $ respectively. From the above, it is not difficult to get $  \forall y:  \eta =  | \vec{t}_{y}|,  $ thus,
	\begin{eqnarray}
	\eta =\frac{24\mathcal{A}_{AB}-12}{ | \vec{s}_{0}| \vec{\hat{s}}_{0}\cdot\vec{\hat{t}}_{0} +  | \vec{s}_{1}| \vec{\hat{s}}_{1}\cdot\vec{\hat{t}}_{1} + | \vec{s}_{2}|  \vec{\hat{s}}_{2}\cdot\vec{\hat{t}}_{2}}
	\label{eq36},
	\end{eqnarray}
	when
	$| \vec{s}_{0}| =  | \vec{s}_{1}|= | \vec{s}_{2}|=\frac{4}{\sqrt{3}}$ and $\vec{\hat{s}}_{0}\cdot\vec{\hat{t}}_{0}=\vec{\hat{s}}_{1}\cdot\vec{\hat{t}}_{1}=\vec{\hat{s}}_{2}\cdot\vec{\hat{t}}_{2}=1$,  we can maximize the denominator to get the lower bound of $\eta$.  Therefore, the lower bound of $ \eta $ is
	\begin{eqnarray}
	\eta\geq\sqrt{3}(2\mathcal{A}_{AB}-1).
	\label{eq37}
	\end{eqnarray}
	This lower bound is nontrivial whenever $ \mathcal{A}_{AB}>\frac{1}{2}$. Next, we consider the witness $ \mathcal{A}_{AC} $. Rewriting inequality (\ref{eq32}), we can get the upper bound of $ \eta$. we show that this upper bound reads
	\begin{eqnarray}
	\eta\leq\frac{1}{2}\sqrt{3(6\sqrt{3}\mathcal{A}_{AC}-3\sqrt{3}+1)(-2\sqrt{3}\mathcal{A}_{AC}+\sqrt{3}+1)}.\label{eq38}
	\end{eqnarray}
	This lower bound is nontrivial whenever $ \frac{1}{2}(1+\frac{\sqrt{3}}{9}) \leq\mathcal{A}_{AC}\leq\frac{1}{2}(1+\frac{\sqrt{3}}{3})$.
	Notice that the upper bound (\ref{eq38}) coincides with the lower bound (\ref{eq37}) for optimal trade-off between the pair  of the two correlation witnesses  ($\mathcal{A}_{AB}$,  $\mathcal{A}_{AC}$) as given in equation (\ref{eq10}).

	In order to have a more intuitive feeling of the noise-tolerance of this scheme in real experiments, we take a simple example to illustrate. When $ \eta=\frac{1}{\sqrt{3}}$,  we attempt to  implement the quantum strategies (\ref{eq31}) and (\ref{eq32}) for the optimal correlation witness pair ($\mathcal{A}_{AB}$,  $\mathcal{A}_{AC}$). In an ideal case, we can easily know ($\mathcal{A}_{AB}$,  $\mathcal{A}_{AC}$) $\approx$ (0.6667, 0.7534).  However, noise is unavoidable in real experiments, and here we can take a $ 95\% $
	visibility in Alice’s preparations, $90\%$ visibility in Bob’s instruments, and $95\%$ visibility in Charlie’s measurements. In this case, we obtain ($\mathcal{A}_{AB}$,  $\mathcal{A}_{AC}$) $\approx$ (0.6425, 0.7156). Therefore, we find  $\eta \in
	[0.4936,0.7844]$. This interval is fairly wide. Note  that the certification is more precise (the interval is smaller) as the sharpness
	parameter increases.    
	It is worth mentioning that the experimental demonstrations for  a three-party prepare-transform-measure protocol with $2 \rightarrow 1 $ sequential QRACs  has been completed in Ref.\cite{anwer2020experimental, foletto2020experimental}. Therefore, we can set the corresponding  experimental scheme of our works by referring to the relevant experimental parameters in the three-party prepare-transform-measure experiment with $2 \rightarrow 1 $ sequential QRACs. This work will be studied in the future.
	\section{Random number generation efficiency in the three-party prepare-transform-measure experiment}
	Before analyzing the randomness of  a three-party prepare-transform-measure experiment,  it should be emphasized firstly, that an optimal weak measurement (that is, the most information can be obtained with the least disturbance to the original quantum state of the system. See  Ref.\cite{silva2015multiple,hu2018observation,an2018experimental} for more details)  is mathematically
	equivalent to POVMs
	formalism \cite{mal2016sharing} and this is the  basis of the  realistic  experiments.  In our scenario, the measurement quality factor $F=\sqrt{1-\eta^{2}}$ and the precision of the measurement $G=\eta$  correspond to the definitions in Ref.\cite{mal2016sharing}. Furthermore, we have $F^{2}+G^{2}=1$. Thus, unsharp measurement yields the maximum information about the system while disturbing the original state minimally. Although  Li  $ et$  $al$.\cite{li2018three} have analyzed the randomness of the classic dimension witnesses (based on the $ 2 \rightarrow 1$ QRAC and the nonlinear determinant value respectively) violation  in the three-observer protocol  by using non-optimal weak measurements,  our analysis is different from theirs because we use  unsharp measurements, which  are  mathematically equivalent to the optimal weak measurement.

According to the observed probabilities,	dimension witness inequality $W$  based on the nonlinear determinant value test in the two-observer system is given by \cite{bowles2014certifying, li2018three,an2018experimental}
	\begin{equation}
W=	\left|\begin{array}{cc}
	p(1|00,0)-p(1|01,0) & p(1|10,0)-p(1|11,0)\\
	p(1|00,1)-p(1|01,1) & p(1|10,1)-p(1|11,1)
	\end{array}\right|.
	\end{equation}
	Specifically, in the	two-dimensional Hilbert space, the upper bound of the quantum	dimension witness value is 1, while the classical dimension
	witness value is 0. This dimension witness can be used to estimate the genuine randomness generated if the quantum random number generator system satisfied two assumptions \cite{bowles2014certifying,lunghi2015self,an2018experimental}. (1) The state
	preparation device and measurement device are assumed to be independent (they have no shared randomness), and their hidden variables are independent of any other devices. For three-party protocol, Alice, Bob, and
	Charlie must be independent with each other.  (2) The dimension of the quantum system is restricted to two.
	
\begin{figure}
		\centering
		\includegraphics[width=3in]{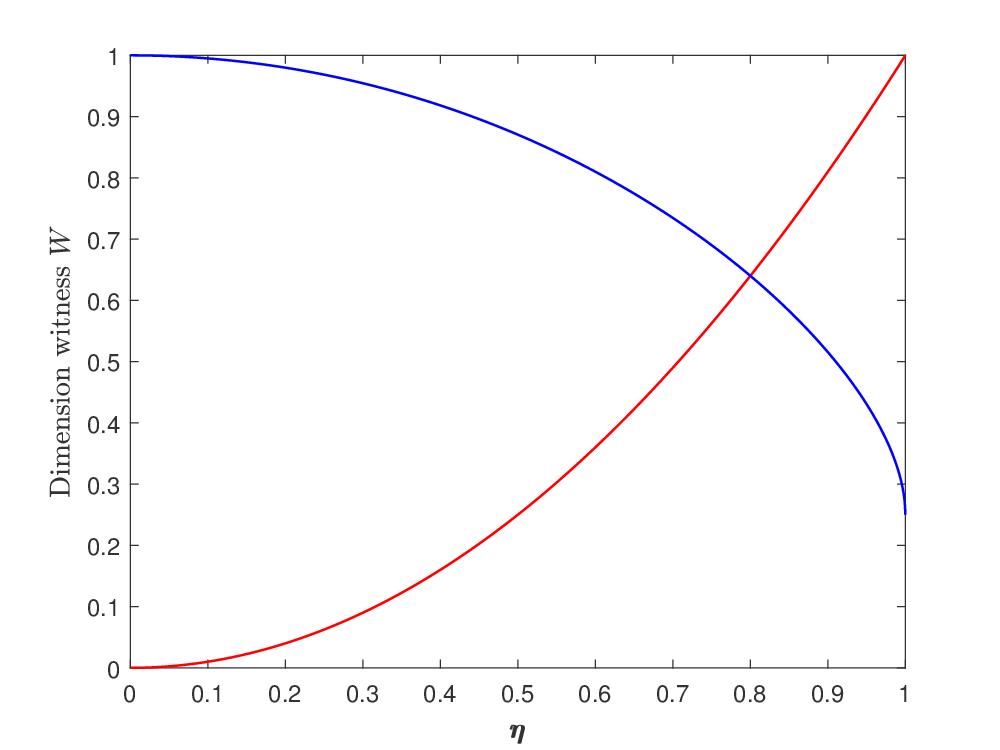}
		\caption{Dimension witness $W$ (based on the determinant value)  as a function
			of the sharpness parameter $ \eta $.
			The blue line is the dimension witness	between Alice and Bob. The red line is  the dimension witness	between Alice and Charlie.}
		\label{fig:4}       
	\end{figure}

	In our scheme, we can obtain the values of the dimension witnesses (based on the determinant value) between Alice and Bob as follows
	\begin{eqnarray}
	W_{AB}=\eta^{2},
	\end{eqnarray}
	while the dimension witness value (based on the determinant value) between Alice and Charlie is given by
	\begin{eqnarray}
	W_{AC}=(\frac{1+\sqrt{1-\eta^{2}}}{2})^{2}.
	\end{eqnarray}
Meanwhile, we find that the sharpness parameters $ \eta $ in this paper are mathematically equivalent to the optimal weak measurement parameters $ \theta $ in Ref.\cite{an2018experimental}, and they satisfy
	\begin{eqnarray}
	\eta=\cos 2\theta.
	\end{eqnarray}
	The detailed quantum dimension witnesses values $ W_{AB}$ and
	$ W_{AC}$ with different sharpness parameters $ \eta $ are shown  in
	Fig.\ref{fig:4}.

	To get more randomness than in Ref.\cite{an2018experimental}, we use the tighter bound of quantum randomness certification given in Ref.\cite{fei2017tighter}, the relation between the dimension witness $W$ (based on the determinant value) and the randomness generation efficiency $ H_{min}$ is given by
	\begin{eqnarray}
	H_{min}(W)=-\log_{2}(\frac{1}{2}+\frac{1}{2}\sqrt{\frac{2-W}{2}}),
	\label{eq42}
	\end{eqnarray}
	where $ 0\leq W\leq 1$. Therefore, we can take the dimensional witnesses values of $ W_{AB}$ and $ W_{AC}$ into equation (\ref{eq42}), and then we can get the local randomness between  on Bob'side and Charlie'side  respectively.

	As is well-known, another quantum dimension witness inequality based on
	the QRAC in the two-observer system is
	given in Ref.\cite{li2011semi,li2012semi}. And we also know that the relation between dimensional witnesses and QRACs (see Ref.\cite{pawlowski2011semi, wehner2008lower}).  Therefore, according to the value of $ \mathcal{A}_{AB}$ and $ \mathcal{A}_{AC}$,
	we  obtain the two-dimensional quantum dimensional witnesses between Alice-Bob and between Alice-Charlie based on $ 2 \rightarrow 1$ QRAC and $ 3 \rightarrow 1$ QRAC respectively, as follows

	\begin{eqnarray}
	T_{AB}^{2 \rightarrow 1}=2\sqrt{2}\eta, \cr
	T_{AC}^{2 \rightarrow 1}=\sqrt{2}(1+\sqrt{1-\eta^{2}}),
	\end{eqnarray}
	and
	\begin{eqnarray}
	T_{AB}^{3 \rightarrow 1}=4\sqrt{3}\eta ,\cr
	T_{AC}^{3\rightarrow 1}=\frac{4\sqrt{3}}{3}(1+2\sqrt{1-\eta^{2}}),
	\end{eqnarray}
	where   $2\leq T^{2 \rightarrow 1}\leq 2\sqrt{2}$ and  $6\leq T^{3 \rightarrow 1}\leq 4\sqrt{3}$.

	In  Ref.\cite{li2015detection}, the general analytical relationship between  dimension witnesses $T^{2 \rightarrow 1}$ and the randomness generation efficiency $ H_{min}^{'}$ is given by

	\begin{eqnarray}
	H_{min}^{'}(T^{2 \rightarrow 1})=-\log_{2}(\frac{1}{2}+\frac{1}{2}\sqrt{\frac{1+\sqrt{1-(\frac{(T^{2 \rightarrow 1})^{2}-4}{4})^{2}}}{2}}).
	\label{eq45}
	\end{eqnarray}
	\begin{figure}
		\centering
		\includegraphics[width=3in]{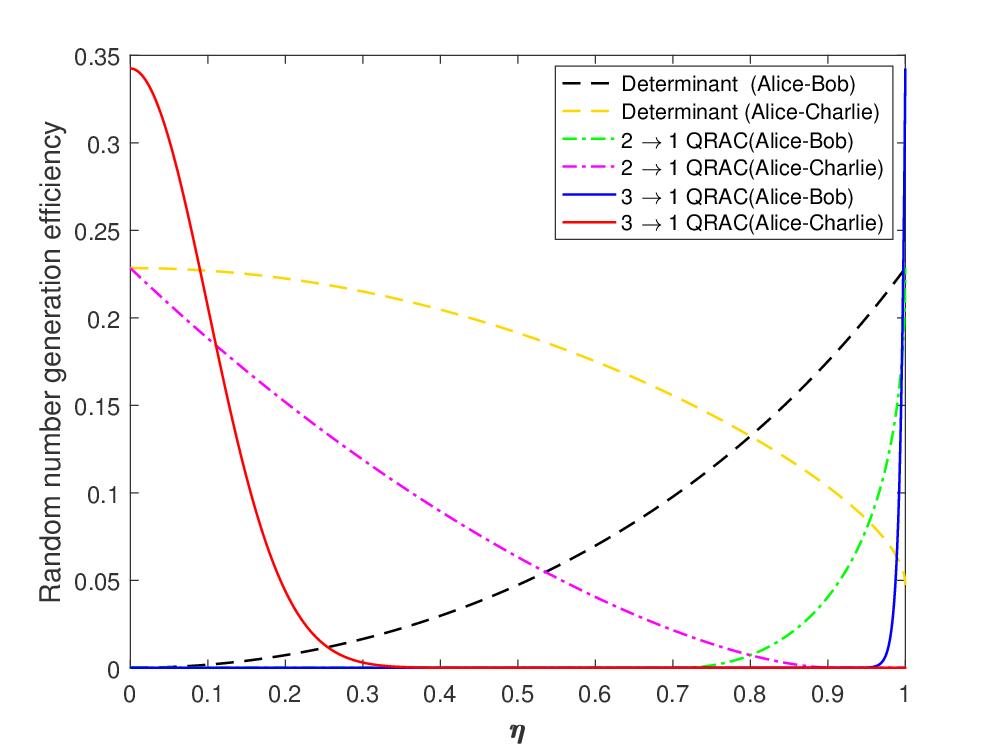}
		\caption{Local random number generation efficiency with different  sharpness parameters $ \eta $.
		}
		\label{fig5:5}       
	\end{figure}
	Hence, we can take the dimensional witnesses values of $ T_{AB}^{2 \rightarrow 1}$ and $ T_{AC}^{2 \rightarrow 1}$ into equation (\ref{eq45}), and then we can get the local randomness on Bob's side and Charlie's side  respectively.
	
	However, in the SDI scenario based on $ 3 \rightarrow 1 $ QRACs, although  the relation between the  randomness generation and the dimension witnesses is also given by using  analytic \cite{zhou2016semi} and numerical analysis  \cite{li2011semi,mironowicz2014properties,navascues2015bounding} respectively,
	we find that the existing analytic relation is not better than the numerical relation.
	Here, we use the numerical method given in  Ref.\cite{li2011semi}. By solving the minimization problem with the Levenberg-
	Marquardt algorithm \cite{levenberg1944method}, we get the min-entropy bound of the measurement outcome for the given  $ 3 \rightarrow 1$  QRAC.  Local random number generation efficiency with different  sharpness parameters $ \eta $ on Bob's side and Charlie's side based on dimensional witnesses values, $ 2 \rightarrow 1$ QRAC and $ 3 \rightarrow 1$ QRAC  are shown in Fig.\ref{fig5:5} respectively. 
	Obviously, if the sharpness parameter is equal to 0, the random generation efficiency is 0.3425, which is same as the result in the SDI randomness expansion with sharp measurements  proposed in Ref. \cite{li2012semi}. It should be noted that in Fig.\ref{fig5:5},
	 the random generation efficiency of $ 3 \rightarrow 1$ QRAC is  better than the results of $ 2 \rightarrow 1$ QRAC on Bob's side (Charlie's side) only when the sharpness parameters  $0.9956<\eta\leq1$  ($0\leq\eta<0.1105$).
	 In other words, the random generation efficiency of $ 3 \rightarrow 1$ QRAC is not always better than the results of $ 2 \rightarrow 1$ QRAC, probably because the tighter  analytic relation (similar to  equation (\ref{eq45})) between random generation efficiency of $ 3 \rightarrow 1$ QRAC and the dimension witness  is missing.
	 
	 The research on the random generation efficiency here is of fundamental significance, and it sheds new light on generating random number among multi-user in the network environment. This also supplies much space to imagine its application in the area of quantum cryptography and quantum randomness generation research.
	\section{Conclusion }
	In conclusion, 
	we derive the optimal  trade-off between the pair  of  two correlation witnesses  ($\mathcal{A}_{AB}$,  $\mathcal{A}_{AC}$) based on $ 3 \rightarrow 1 $ sequential QRACs  in a three-party prepare-transform-measure experiment.  Based on the    trade-off, we have completed the self-testing  of preparations,  instruments and measurements  for three sequential  parties.
	We give the upper and lower bounds of the sharpness parameter, and complete the robustness analysis of the self-testing scheme.
	We find that  classical correlation witness violation based on $3 \rightarrow 1 $ sequential RACs  cannot be obtained  by both  correlation witnesses $\mathcal{A}_{AB}$ and   $\mathcal{A}_{AC}$ simultaneously. This implies that if Bob  uses unsharp measurements strong enough to
	achieve
	the maximal classical  violation of the correlation witness, Charlie cannot do so even with maximal strength.
	Besides,  we   give the  analysis and comparison  of the  random number generation efficiency under different sharpness parameters based on the  determinant value,  $2 \rightarrow 1 $ and  $3 \rightarrow 1 $ QRACs  respectively, and the analysis method can also be applied to future multi-party quantum network studies.
	
	\section*{Acknowledegments}
	This work is supported by the  National Natural Science Foundation of China (Grant Nos. 61672110, 61671082, 61976024, 61972048), and the Fundamental Research Funds for the Central Universities (Grant No.2019XD-A01).
	
	\section*{Appendix A. Proof of Proposition 1 }
	
	\setcounter{equation}{0}
	
	\renewcommand\theequation{A\arabic{equation}}
	
	In this section  we provide the proof of Proposition 1, which for completeness we also state here.
	
		$\mathbf{Proposition}$ $\mathbf{1}$. The optimal trade-off between the pair  of the two correlation witnesses  ($\mathcal{A}_{AB}$,  $\mathcal{A}_{AC}$) based on $ 3 \rightarrow 1 $ sequential QRACs corresponds to
		\begin{eqnarray}
		\mathcal{A}_{AC}^{\mathcal{A}_{AB}}=\frac{1}{2}+\frac{\sqrt{3}}{18}(1+2\sqrt{12\mathcal{A}_{AB}-12\mathcal{A}_{AB}^{2}-2}),
		\label{eq10}
		\end{eqnarray}
		where $ \mathcal{A}_{AB}\in[1/2,(1+1/\sqrt{3})/2] $.
		
		$\mathbf{Proof}$: We  use the polar decomposition to write the Kraus operators as
		$K_{b\mid y} = U_{yb}\sqrt{B_{b| y}} $ for the unitary operator $U_{yb}$ and the  element of POVM {$B_{b| y}$} in the above. Kraus operators of this form  correspond to extremal quantum instruments in the considered scenario [18]. We can then use the cyclicity of the trace along with the substitution $C_{1\mid z}=I-C_{0\mid z} $ to write
		equation (\ref{eq8}) as
		\begin{eqnarray}
		\mathcal{A}_{AC}= \frac{1}{2}+\frac{1}{72}\sum_{x,y,b,z}(-1)^{x_{z}}  \mathrm{tr} [\sqrt{B_{b| y}}\rho_{x}\sqrt{B_{b| y}}U_{yb}^{\dagger}C_{0| z}U_{yb}]\cr
		= \frac{1}{2}+\frac{1}{72}\sum_{x,y,b,z}\mathrm{tr} [\sqrt{B_{b| y}}\gamma_{z}\sqrt{B_{b| y}}V_{yzb}],
		\end{eqnarray}
		where $\gamma_{z}=\sum_{x}(-1)^{x_{z}}\rho_{x}    $
		, $ V_{yzb}=U_{yb}^{\dagger}C_{0| z}U_{yb} $.
		We can now consider the optimisation over {$ U_{yb}$} and $C_{c\mid z}$ as a single optimisation over $V_{yzb}$. To this end, we note that the set of measurements \{$C_{c\mid z}$\} is convex. Therefore, every nonextremal (interior
		point) measurement can be written as a convex combination of extremal measurements (on the boundary). Due to linearity, no nonextremal POVM can lead to a larger value of $ \mathcal{A}_{AC} $ than some extremal POVM. The extremal binary-outcome qubit measurements are rank-one projectors. Therefore, we can consider the optimisation over $ V_{yzb}$  as an optimisation over general rank-one projectors. This gives
		\begin{eqnarray}
		\max   \mathcal{A}_{AC}=\frac{1}{2}+\max_{\rho,V,B}\frac{1}{72}\sum_{y,b,z}\mathrm{tr} [\sqrt{B_{b| y}}\gamma_{z}\sqrt{B_{b| y}}V_{zyb}]\cr
		=\frac{1}{2}+ \max_{\rho,B}\frac{1}{72}\sum_{y,b,z}\lambda_{\max}[\sqrt{B_{b| y}}\gamma_{z}\sqrt{B_{b| y}}],
		\label{eq12}
		\end{eqnarray}
		where  the optimal choice of  $ V_{yzb} $  is aligned with the eigenvector of $\sqrt{B_{b| y}}\gamma_{z}\sqrt{B_{b| y}} $ corresponding to the largest eigenvalue $ \lambda_{\max} $.
		
		The general representation of a qubit can be illustrated by using the density  matrix formalism $\rho_{x}=\frac{I+\vec{n}_{x}\cdot\vec{\sigma}}{2}, $ Bloch vector $ \vec{n}_{x}\in R^{3},$ $| \vec{n}_{x}| \leq 1$. Thus, we have
		

		
		\begin{eqnarray}
		\gamma_{0}=\frac{1}{2}[\vec{n}_{000}+\vec{n}_{001}+\vec{n}_{010}+\vec{n}_{011}-(\vec{n}_{111}+\vec{n}_{110}+\vec{n}_{101}+\vec{n}_{100})]\cdot\vec{\sigma}=\vec{s}_{0}\cdot\vec{\sigma}, \cr
		\gamma_{1}=\frac{1}{2}[\vec{n}_{000}+\vec{n}_{001}+\vec{n}_{101}+\vec{n}_{100}-(\vec{n}_{010}+\vec{n}_{011}+\vec{n}_{111}+\vec{n}_{110})]\cdot\vec{\sigma}=\vec{s}_{1}\cdot\vec{\sigma}, \cr
		\gamma_{2}=\frac{1}{2}[\vec{n}_{000}+\vec{n}_{010}+\vec{n}_{110}+\vec{n}_{100}-(\vec{n}_{001}+\vec{n}_{011}+\vec{n}_{111}+\vec{n}_{101})]\cdot\vec{\sigma}=\vec{s}_{2}\cdot\vec{\sigma}, \cr
		\end{eqnarray}
		where
		\begin{eqnarray}
		\vec{s}_{0}=\frac{1}{2}[\vec{n}_{000}+\vec{n}_{001}+\vec{n}_{010}+\vec{n}_{011}-(\vec{n}_{111}+\vec{n}_{110}+\vec{n}_{101}+\vec{n}_{100})],\cr\label{eq14}
		\vec{s}_{1}=\frac{1}{2}[\vec{n}_{000}+\vec{n}_{001}+\vec{n}_{101}+\vec{n}_{100}-(\vec{n}_{010}+\vec{n}_{011}+\vec{n}_{111}+\vec{n}_{110})],\cr
		\vec{s}_{2}=\frac{1}{2}[\vec{n}_{000}+\vec{n}_{010}+\vec{n}_{110}+\vec{n}_{100}-(\vec{n}_{001}+\vec{n}_{011}+\vec{n}_{111}+\vec{n}_{101})].
		\end{eqnarray}
		
		As we know, given any set of preparations $\{\vec{n}_{x}\}$, we can consider other preparations $\{\vec{\tilde{n}}_{x}\} $ choosen such that $\vec{\vec{\tilde{n}}}_{000}=-\vec{\tilde{n}}_{111} $, $\vec{\tilde{n}}_{001}=-\vec{\tilde{n}}_{110} $, $\vec{\tilde{n}}_{010}=-\vec{\tilde{n}}_{101} $, $\vec{\tilde{n}}_{011}=-\vec{\tilde{n}}_{100} $.
		Therefore, we can reduce the number of operator equalities (\ref{eq14}) by
		exploiting the apparent symmetries in the expressions for $\vec{s}_{z}$. 
		Moreover, it is evident that if not all preparations are pure, one cannot obtain optimal correlations. Thus, without loss of generality, we define
		\begin{eqnarray}
		\vec{n}_{000}=(\sin \mu\cos \varphi,\sin \mu\sin \varphi,\cos\mu),\cr
		\vec{n}_{001}=(\sin \mu\cos \varphi,\sin \mu\sin \varphi,-\cos\mu),\cr
		\vec{n}_{010}=(\sin \mu\cos \varphi,-\sin \mu\sin \varphi,\cos\mu),\cr
		\vec{n}_{011}=(\sin \mu\cos \varphi,-\sin \mu\sin \varphi,-\cos\mu),
		\end{eqnarray}
		where $ \mu,\varphi\in[0,\frac{\pi}{2}].$
		This lead to
		\begin{eqnarray}
		| \vec{s}_{0}| =4\sin \mu\cos \varphi , | \vec{s}_{1}| =4\sin \mu\sin \varphi , |\vec{s}_{2}| =4\cos\mu.
		\end{eqnarray}
		
		To further derive the upper bound of equation (\ref{eq12}), we can use the following relation
		\begin{eqnarray}
		\forall B,\forall\vec{a}\in R^{3}:\sum_{b=0,1}\lambda_{\max}[\sqrt{B_{b}}(\vec{a}\cdot\vec{\sigma})\sqrt{B_{b}}]\leq| \vec{a}|,
		\end{eqnarray}
		with equality if and only if $ \vec{a}$ is aligned with the Bloch vector of the POVM. Identifying $\vec{a}$ with $\vec{s}_{z}$, we apply it
		twice to equation (12) corresponding to the terms in which $ z = y$. This gives
		\begin{eqnarray}
		\mathcal{A}_{AC}\leq \frac{1}{2}+\frac{1}{72}( | \vec{s}_{0}| + | \vec{s}_{1}| +| \vec{s}_{2}|+\sum_{y,b}\lambda_{\max}[\sqrt{B_{b| y}}(\vec{s}_{\tilde{y}}\cdot\vec{\sigma})\sqrt{B_{b| y}}]),\label{eq18}
		\end{eqnarray}
		where $y,\tilde{y}\in\{0,1,2\}  $ and $y\neq\tilde{y}$.

		For the convenience of the following analysis,  we define
		$ B_{y}=\alpha_{y}I+\vec{t}_{y}\cdot\vec{\sigma} $ where $\vec{t}_{y}=(t_{y0}$, $t_{y1},t_{y2}),$ $| \vec{t}_{y}| \leq 1,$ $ | \vec{t}_{y}|-1\leq \alpha_{y}\leq 1-| \vec{t}_{y}|$. The sharpness parameter of Bob’s measurements is defined as $ \eta_{y}= | \vec{t}_{y}| $. Notice that for $ \eta_{y}\in\{0,1\}$  the measurements are noninteractive and sharp  measurement respectively,
		whereas $ \eta_{y}\in(0,1)$ corresponds to intermediate cases.
		Furthermore, we have
		\begin{eqnarray}
		B_{b| y}=f_{yb}| \vec{t}_{y} \rangle\langle\vec{t}_{y}| + h_{yb}|-\vec{t}_{y} \rangle\langle-\vec{t}_{y}|,\label{eq19}
		\end{eqnarray}
		where $ | \vec{t}_{y} \rangle$ is  the pure state corresponding to the Bloch sphere direction  $\vec{t}_{y}$, and
		
		\begin{eqnarray}
		f_{yb}=\frac{1}{2}(1+(-1)^{b}\alpha_{y}+(-1)^{b}|\vec{t}_{y}|),\cr
		h_{yb}=\frac{1}{2}(1+(-1)^{b}\alpha_{y}-(-1)^{b}|\vec{t}_{y}|).\label{eq20}
		\end{eqnarray}
		
		From the above analysis, the equation  (\ref{eq4}) can be written as
		\begin{eqnarray}
		\mathcal{A}_{AB}=\frac{1}{2}+\frac{1}{24}(\vec{s}_{0}\cdot\vec{t}_{0}+\vec{s}_{1}\cdot\vec{t}_{1}+\vec{s}_{2}\cdot\vec{t}_{2})\cr
		=\frac{1}{2}+\frac{1}{24}( | \vec{s}_{0}| t_{00} + | \vec{s}_{1}| t_{11} +|\vec{s}_{2}| t_{22}).\label{eq21}
		\end{eqnarray}
		
		Then, in order to derive the upper bound of $\mathcal{A}_{AC}$, we need to further analyze inequality (\ref{eq18}), consider the characteristic equation det $ (\sqrt{B_{b| y}}(\vec{s}_{\tilde{y}}\cdot\vec{\sigma})\sqrt{B_{b| y}}-\mu I)=0  $
		and after a complicated derivation and simplification, we can obtain
		\begin{eqnarray}
		T=\sum_{y,b}\lambda_{\max}[\sqrt{B_{b| y}}(\vec{s}_{\tilde{y}}\cdot\vec{\sigma})\sqrt{B_{b| y}}]\cr
		=\sum_{y,b}\frac{| \vec{s}_{\tilde{y}}|}{2}\sqrt{(1+(-1)^{b}\alpha_{y})^{2}-|\vec{t}_{y}|^{2}(1-\langle\vec{t}_{y}| \vec{\hat{s}}_{\tilde{y}}| \vec{t}_{y} \rangle^{2})},\label{eq22}
		\end{eqnarray}
		where $\vec{\hat{s}}$ is the normalized form of $ \vec{s} $.
		We can now consider the optimisation over $ \alpha_{y}$ by separately considering the three terms corresponding to $ y = 0$,   $y = 1$ and $y=2$ respectively. This amounts to maximising expressions of the form $\sqrt{(1 + x)^2-R} +\sqrt{(1 - x)^2-R} $, for some positive constant $ R $. It is easily shown that the value of function is maximal if and only if  $x = 0$. Thus, we require $\alpha_{0}  = \alpha_{1}=\alpha_{2}   = 0.$ Moreover, since $\vec{s}_{0}\propto(1,0,0)$, $\vec{s}_{1}\propto(0,1,0) $ and $\vec{s}_{2}\propto(0,0,1)$, and we also separately maximise search squareroot
		expression above by standard differentiation, it is seen from (\ref{eq21}) and (\ref{eq22}) that one optimally chooses $ t_{01}=t_{02}=0$, $t_{10}=t_{12}=0$, $t_{20}=t_{21}=0 $. Finally, we can get
		\begin{eqnarray}
		\max T=|\vec{s}_{0}|(\sqrt{1-t_{11}^{2}}+\sqrt{1-t_{22}^{2}})+|\vec{s}_{1}|(\sqrt{1-t_{00}^{2}}+\sqrt{1-t_{22}^{2}})\cr
		+|\vec{s}_{2}|(\sqrt{1-t_{00}^{2}}+\sqrt{1-t_{11}^{2}}).
		\end{eqnarray}
		Inequality (\ref{eq18}) can be written as
		\begin{eqnarray}
		\mathcal{A}_{AC}\leq \frac{1}{2}+\frac{1}{72}( | \vec{s}_{0}| + | \vec{s}_{1}| +| \vec{s}_{2}|+|\vec{s}_{0}|(\sqrt{1-t_{11}^{2}}
		+\sqrt{1-t_{22}^{2}})\cr
		+|\vec{s}_{1}|(\sqrt{1-t_{00}^{2}}+\sqrt{1-t_{22}^{2}})+|
		\vec{s}_{2}|(\sqrt{1-t_{00}^{2}}+\sqrt{1-t_{11}^{2}}))\equiv\mathcal{A}_{C}.\label{eq24}
		\end{eqnarray}
		Without loss of generality, we define  $t_{00}=\cos \phi_{0}$, $t_{11}=\cos \phi_{1},$ $ t_{22}=\cos \phi_{2} $ where $  \phi_{y}\in[0,$ $\frac{\pi}{2}]$. By plugging this in inequality (\ref{eq24}), $\mathcal{A}_{C}$ can can be reexpressed
		\begin{eqnarray}
		\mathcal{A}_{C}=\frac{1}{2}+\frac{1}{18}[\sin \mu\cos \varphi+
		\sin \mu\sin \varphi+\cos \mu+
		\sin \mu\cos \varphi(\sin \phi_{1}\cr
		+\sin \phi_{2})+\sin \mu\sin \varphi(\sin \phi_{0}+\sin \phi_{2})+\cos \mu(\sin \phi_{0}+\sin \phi_{1})].\label{eq25}
		\end{eqnarray}
		Accordingly, equation (\ref{eq21}) can be rewritten as
		\begin{eqnarray}
		\mathcal{A}_{AB}=\frac{1}{2}+\frac{1}{6}(\sin \mu\cos \varphi\cos \phi_{0}+\sin \mu\sin \varphi\cos \phi_{1}+\cos \mu\cos \phi_{2}).\label{eq26}
		\end{eqnarray}
		Next, the optimization problem is transformed into taking the values of all parameters   $  \mu,$ $\varphi,$ $\phi_{0},$ $\phi_{1},\phi_{2} $, and finding the maximum value of $\mathcal{A}_{C}$. To solve this problem, we use the following theorem
		
		$\mathbf{Lemma}$ $\mathbf{1}$. For every tuple ($  \mu,$ $\varphi,$ $\phi_{0},$ $\phi_{1},\phi_{2} $) corresponding to ($\mathcal{A}_{AB},\mathcal{A}_{C} $), there exists another tuple $ (  \mu,$ $\varphi,$ $\phi_{0},$ $\phi_{1},\phi_{2} )=(\arccos \frac{1}{\sqrt{3}},$ $\frac{\pi}{4},$ $\phi,$ $\phi,$ $\phi )$ that always produces ($\mathcal{A}_{AB},  \tilde{\mathcal{A}}_{C} $) with $\tilde{\mathcal{A}}_{C}\geq \mathcal{A}_{C}$.

		To prove this statement, we must show that for all $  \mu,\varphi,\phi_{0},\phi_{1},\phi_{2}\in[0,\frac{\pi}{2}] $ there exists a $\phi\in[0,\frac{\pi}{2}]$ such that
		\begin{eqnarray}
		\sin \mu\cos \varphi\cos \phi_{0}+\sin \mu\sin \varphi\cos \phi_{1}+\cos \mu\cos \phi_{2}=\sqrt{3}\cos\phi,\cr
		\sin \mu\cos \varphi(\sin \phi_{1}
		+\sin \phi_{2})+\sin \mu\sin \varphi(\sin \phi_{0}+\sin \phi_{2})\cr
		+\cos \mu(\sin \phi_{0}+\sin \phi_{1})
		\leq  \sqrt{3}+2\sqrt{3}\sin \phi.
		\label{eq27}
		\end{eqnarray}
		As far as we know, it trivially holds that $\sin \mu\cos \varphi+\sin \mu\sin \varphi+\cos \mu \leq  \sqrt{3}$ with equality if and only if $\mu= \arccos \frac{1}{\sqrt{3}},\varphi=\frac{\pi}{4} $.
		Furthermore, we  obtain
		\begin{eqnarray}
		\sin \mu\cos \varphi \sin \phi_{1}+\sin \mu\sin \varphi \sin \phi_{2}+\cos \mu\sin \phi_{0}\leq \sqrt{3}\sin\phi,\cr\label{eq28}
		\sin \mu\cos \varphi \sin \phi_{2}+\sin \mu\sin \varphi \sin \phi_{0}+\cos \mu\sin \phi_{1}\leq \sqrt{3}\sin\phi,\cr
		\sin \mu\cos \varphi \cos \phi_{0}+\sin \mu\sin \varphi \cos \phi_{1}+\cos \mu\cos \phi_{2}\leq \sqrt{3}\cos\phi.
		\end{eqnarray}
		Then, by squaring  inequations  (\ref{eq27}) and inequations  (\ref{eq28}), we have
		\begin{eqnarray}
		\frac{1}{3}(\sin ^{2}\phi_{1}+\sin ^{2}\phi_{2}+\sin ^{2}\phi_{0})+\frac{2}{3}(\sin \phi_{1}\sin \phi_{2}+\sin \phi_{0}\sin \phi_{1}+\sin \phi_{0}\sin \phi_{2})\leq  3\sin ^{2}\phi,\cr \label{eq29}
		\frac{1}{3}(\sin ^{2}\phi_{0}+\sin ^{2}\phi_{1}+\sin ^{2}\phi_{2})+\frac{2}{3}(\sin \phi_{0}\sin \phi_{2}+\sin \phi_{1}\sin \phi_{2}+\sin \phi_{0}\sin \phi_{1})\leq  3\sin ^{2}\phi,\cr
		\frac{1}{3}(\cos ^{2}\phi_{0}+\cos ^{2}\phi_{1}+\cos ^{2}\phi_{2})+\frac{2}{3}(\cos \phi_{1}\cos \phi_{2}+\cos \phi_{0}\cos \phi_{1}+\cos \phi_{0}\cos \phi_{2})\leq  3\cos ^{2}\phi. \cr
		\end{eqnarray}
		we can combine inequations (\ref{eq29}) into a single equation in which $\phi$  is eliminated. The statement reduces to the inequality
		\begin{eqnarray}
		\sin \phi_{1}\sin \phi_{2}+\sin \phi_{0}\sin \phi_{1}+\sin \phi_{0}\sin \phi_{2}\cr
		+\cos \phi_{1}\cos \phi_{2}+\cos \phi_{0}\cos \phi_{1}+\cos \phi_{0}\cos \phi_{2}\leq 3.
		\end{eqnarray}
		After a detailed derivation, one finds that the optimum of the left hand side is attained for $\phi_{0}=\phi_{1}=\phi_{2}$. Then, the theorem 1 can be derived.
		
		Putting these together we can reduce our consideration of (\ref{eq25}) and (\ref{eq26}) to $ \mu= \arccos \frac{1}{\sqrt{3}},
		\varphi=\frac{\pi}{4},$ $t_{00}=\cos \phi_{0}=\eta_{0},$ $t_{11}=\cos \phi_{1}=\eta_{1},$ $t_{22}=\cos \phi_{2}=\eta_{2}$ and $ \eta_{0}=\eta_{1}=\eta_{2}\equiv \eta $.
		Therefore, equation (\ref{eq26}) reduces to
		\begin{eqnarray}
		\mathcal{A}_{AB}=\frac{1}{2}+\frac{\sqrt{3}}{6}\eta,\label{eq31}
		\end{eqnarray}
		and equation (\ref{eq25}) reduces to
		\begin{eqnarray}
		\mathcal{A}_{AC}=\frac{1}{2}+\frac{\sqrt{3}}{18}(1+2\sqrt{1-\eta^2}).\label{eq32}
		\end{eqnarray}

		According to equation (\ref{eq31}), we have
		$\eta=\sqrt{3}(2\mathcal{A}_{AB}-1)$.
		By plugging this in equation (\ref{eq32}), we finally get
		\begin{eqnarray}
		\mathcal{A}_{AC}^{\mathcal{A}_{AB}}=\frac{1}{2}+\frac{\sqrt{3}}{18}(1+2\sqrt{12\mathcal{A}_{AB}-12\mathcal{A}_{AB}^{2}-2}).
		\end{eqnarray}
		This finishes the proof of Proposition 1. 
		$\hfill\qedsymbol$


\begin{thebibliography}{10}

\bibitem{heisenberg1949physical}
Werner Heisenberg.
\newblock {\em The physical principles of the quantum theory}.
\newblock Courier Corporation, 1949.

\bibitem{aharonov1988result}
Yakir Aharonov, David~Z Albert, and Lev Vaidman.
\newblock How the result of a measurement of a component of the spin of a
  spin-1/2 particle can turn out to be 100.
\newblock {\em Physical Review Letters}, 60(14):1351, 1988.

\bibitem{curchod2017unbounded}
Florian~J Curchod, Markus Johansson, Remigiusz Augusiak, Matty~J Hoban, Peter
  Wittek, and Antonio Ac{\'\i}n.
\newblock Unbounded randomness certification using sequences of measurements.
\newblock {\em Physical Review A}, 95(2):020102, 2017.

\bibitem{li2018three}
Hong-Wei Li, Yong-Sheng Zhang, Xue-Bi An, Zheng-Fu Han, and Guang-Can Guo.
\newblock Three-observer classical dimension witness violation with weak
  measurement.
\newblock {\em Communications Physics}, 1(1):1--8, 2018.

\bibitem{an2018experimental}
Xue-Bi An, Hong-Wei Li, Zhen-Qiang Yin, Meng-Jun Hu, Wei Huang, Bing-Jie Xu,
  Shuang Wang, Wei Chen, Guang-Can Guo, and Zheng-Fu Han.
\newblock Experimental three-party quantum random number generator based on
  dimension witness violation and weak measurement.
\newblock {\em Optics Letters}, 43(14):3437--3440, 2018.

\bibitem{coyle2018one}
Brian Coyle, Matty~J Hoban, and Elham Kashefi.
\newblock One-sided device-independent certification of unbounded random
  numbers.
\newblock {\em arXiv preprint arXiv:1806.10565}, 2018.

\bibitem{lundeen2012procedure}
Jeff~S Lundeen and Charles Bamber.
\newblock Procedure for direct measurement of general quantum states using weak
  measurement.
\newblock {\em Physical Review Letters}, 108(7):070402, 2012.

\bibitem{wu2013state}
Shengjun Wu.
\newblock State tomography via weak measurements.
\newblock {\em Scientific Reports}, 3:1193, 2013.

\bibitem{silva2015multiple}
Ralph Silva, Nicolas Gisin, Yelena Guryanova, and Sandu Popescu.
\newblock Multiple observers can share the nonlocality of half of an entangled
  pair by using optimal weak measurements.
\newblock {\em Physical Review Letters}, 114(25):250401, 2015.

\bibitem{shenoy2019unbounded}
Akshata Shenoy, S{\'e}bastien Designolle, Flavien Hirsch, Ralph Silva, Nicolas
  Gisin, and Nicolas Brunner.
\newblock Unbounded sequence of observers exhibiting einstein-podolsky-rosen
  steering.
\newblock {\em Physical Review A}, 99(2):022317, 2019.

\bibitem{anwer2019noise}
Hammad Anwer, Natalie Wilson, Ralph Silva, Sadiq Muhammad, Armin Tavakoli, and
  Mohamed Bourennane.
\newblock Noise-robust preparation contextuality shared between any number of
  observers via unsharp measurements.
\newblock {\em arXiv preprint arXiv:1904.09766}, 2019.

\bibitem{brown2020arbitrarily}
Peter~J Brown and Roger Colbeck.
\newblock Arbitrarily many independent observers can share the nonlocality of a
  single maximally entangled qubit pair.
\newblock {\em Physical Review Letters}, 125(9):090401, 2020.

\bibitem{mohan2019sequential}
Karthik Mohan, Armin Tavakoli, and Nicolas Brunner.
\newblock Sequential random access codes and self-testing of quantum
  measurement instruments.
\newblock {\em New Journal of Physics}, 21(8):083034, 2019.

\bibitem{ambainis1999dense}
Andris Ambainis, Ashwin Nayak, Ammon Ta-Shma, and Umesh Vazirani.
\newblock Dense quantum coding and a lower bound for 1-way quantum automata.
\newblock In {\em Proceedings of the thirty-first annual ACM symposium on
  Theory of computing}, pages 376--383, 1999.

\bibitem{ambainis2008quantum}
Andris Ambainis, Debbie Leung, Laura Mancinska, and Maris Ozols.
\newblock Quantum random access codes with shared randomness.
\newblock {\em arXiv preprint arXiv:0810.2937}, 2008.

\bibitem{anwer2020experimental}
Hammad Anwer, Sadiq Muhammad, Walid Cherifi, Nikolai Miklin, Armin Tavakoli,
  and Mohamed Bourennane.
\newblock Experimental characterization of unsharp qubit observables and
  sequential measurement incompatibility via quantum random access codes.
\newblock {\em Physical Review Letters}, 125(8):080403, 2020.

\bibitem{foletto2020experimental}
Giulio Foletto, Luca Calderaro, Giuseppe Vallone, and Paolo Villoresi.
\newblock Experimental demonstration of sequential quantum random access codes.
\newblock {\em Physical Review Research}, 2(3):033205, 2020.

\bibitem{popescu1992states}
Sandu Popescu and Daniel Rohrlich.
\newblock Which states violate bell's inequality maximally?
\newblock {\em Physics Letters A}, 169(6):411--414, 1992.

\bibitem{clauser1969proposed}
John~F Clauser, Michael~A Horne, Abner Shimony, and Richard~A Holt.
\newblock Proposed experiment to test local hidden-variable theories.
\newblock {\em Physical Review Letters}, 23(15):880, 1969.

\bibitem{yang2013robust}
Tzyh~Haur Yang and Miguel Navascu{\'e}s.
\newblock Robust self-testing of unknown quantum systems into any entangled
  two-qubit states.
\newblock {\em Physical Review A}, 87(5):050102, 2013.

\bibitem{wu2014robust}
Xingyao Wu, Yu~Cai, Tzyh~Haur Yang, Huy~Nguyen Le, Jean-Daniel Bancal, and
  Valerio Scarani.
\newblock Robust self-testing of the three-qubit w state.
\newblock {\em Physical Review A}, 90(4):042339, 2014.

\bibitem{pal2014device}
K{\'a}roly~F P{\'a}l, Tam{\'a}s V{\'e}rtesi, and Miguel Navascu{\'e}s.
\newblock Device-independent tomography of multipartite quantum states.
\newblock {\em Physical Review A}, 90(4):042340, 2014.

\bibitem{kaniewski2017self}
J{\k{e}}drzej Kaniewski.
\newblock Self-testing of binary observables based on commutation.
\newblock {\em Physical Review A}, 95(6):062323, 2017.

\bibitem{coladangelo2017all}
Andrea Coladangelo, Koon~Tong Goh, and Valerio Scarani.
\newblock All pure bipartite entangled states can be self-tested.
\newblock {\em Nature Communications}, 8(1):1--5, 2017.

\bibitem{baccari2020scalable}
F~Baccari, R~Augusiak, I~{\v{S}}upi{\'c}, J~Tura, and A~Ac{\'\i}n.
\newblock Scalable bell inequalities for qubit graph states and robust
  self-testing.
\newblock {\em Physical Review Letters}, 124(2):020402, 2020.

\bibitem{tavakoli2018self}
Armin Tavakoli, J{\k{e}}drzej Kaniewski, Tam{\'a}s V{\'e}rtesi, Denis Rosset,
  and Nicolas Brunner.
\newblock Self-testing quantum states and measurements in the
  prepare-and-measure scenario.
\newblock {\em Physical Review A}, 98(6):062307, 2018.

\bibitem{farkas2019self}
M{\'a}t{\'e} Farkas and J{\k{e}}drzej Kaniewski.
\newblock Self-testing mutually unbiased bases in the prepare-and-measure
  scenario.
\newblock {\em Physical Review A}, 99(3):032316, 2019.

\bibitem{mironowicz2019experimentally}
Piotr Mironowicz and Marcin Paw{\l}owski.
\newblock Experimentally feasible semi-device-independent certification of
  four-outcome positive-operator-valued measurements.
\newblock {\em Physical Review A}, 100(3):030301, 2019.

\bibitem{wei2019robustness}
Shi-Hui Wei, Fen-Zhuo Guo, Xin-Hui Li, and Qiao-Yan Wen.
\newblock Robustness self-testing of states and measurements in the
  prepare-and-measure scenario with random access code.
\newblock {\em Chinese Physics B}, 28(7):070304, 2019.

\bibitem{miklin2020semi}
Nikolai Miklin, Jakub~J Borka{\l}a, and Marcin Paw{\l}owski.
\newblock Semi-device-independent self-testing of unsharp measurements.
\newblock {\em Physical Review Research}, 2(3):033014, 2020.

\bibitem{tavakoli2020self}
Armin Tavakoli, Massimiliano Smania, Tam{\'a}s V{\'e}rtesi, Nicolas Brunner,
  and Mohamed Bourennane.
\newblock Self-testing nonprojective quantum measurements in
  prepare-and-measure experiments.
\newblock {\em Science Advances}, 6(16):eaaw6664, 2020.

\bibitem{miklin2020universal}
Nikolai Miklin and Micha{\l} Oszmaniec.
\newblock A universal scheme for robust self-testing in the prepare-and-measure
  scenario.
\newblock {\em arXiv preprint arXiv:2003.01032}, 2020.

\bibitem{tavakoli2020semi}
Armin Tavakoli.
\newblock Semi-device-independent certification of independent quantum state
  and measurement devices.
\newblock {\em Physical Review Letters}, 125(15):150503, 2020.

\bibitem{mal2016sharing}
Shiladitya Mal, Archan~S Majumdar, and Dipankar Home.
\newblock Sharing of nonlocality of a single member of an entangled pair of
  qubits is not possible by more than two unbiased observers on the other wing.
\newblock {\em Mathematics}, 4(3):48, 2016.

\bibitem{maity2020detection}
Ananda~G Maity, Debarshi Das, Arkaprabha Ghosal, Arup Roy, and AS~Majumdar.
\newblock Detection of genuine tripartite entanglement by multiple sequential
  observers.
\newblock {\em Physical Review A}, 101(4):042340, 2020.

\bibitem{navascues2015bounding}
Miguel Navascu{\'e}s and Tam{\'a}s V{\'e}rtesi.
\newblock Bounding the set of finite dimensional quantum correlations.
\newblock {\em Physical Review Letters}, 115(2):020501, 2015.

\bibitem{li2012semi}
Hong-Wei Li, Marcin Paw{\l}owski, Zhen-Qiang Yin, Guang-Can Guo, and Zheng-Fu
  Han.
\newblock Semi-device-independent randomness certification using n→ 1 quantum
  random access codes.
\newblock {\em Physical Review A}, 85(5):052308, 2012.

\bibitem{hu2018observation}
Meng-Jun Hu, Zhi-Yuan Zhou, Xiao-Min Hu, Chuan-Feng Li, Guang-Can Guo, and
  Yong-Sheng Zhang.
\newblock Observation of non-locality sharing among three observers with one
  entangled pair via optimal weak measurement.
\newblock {\em NPJ Quantum Information}, 4(1):1--7, 2018.

\bibitem{bowles2014certifying}
Joseph Bowles, Marco~T{\'u}lio Quintino, and Nicolas Brunner.
\newblock Certifying the dimension of classical and quantum systems in a
  prepare-and-measure scenario with independent devices.
\newblock {\em Physical Review Letters}, 112(14):140407, 2014.

\bibitem{lunghi2015self}
Tommaso Lunghi, Jonatan~Bohr Brask, Charles Ci~Wen Lim, Quentin Lavigne, Joseph
  Bowles, Anthony Martin, Hugo Zbinden, and Nicolas Brunner.
\newblock Self-testing quantum random number generator.
\newblock {\em Physical Review Letters}, 114(15):150501, 2015.

\bibitem{fei2017tighter}
Xin-Wei Fei, Zhen-Qiang Yin, Wei Huang, Bing-Jie Xu, Shuang Wang, Wei Chen,
  Yun-Guang Han, Guang-Can Guo, and Zheng-Fu Han.
\newblock Tighter bound of quantum randomness certification for
  independent-devices scenario.
\newblock {\em Scientific Reports}, 7(1):1--6, 2017.

\bibitem{li2011semi}
Hong-Wei Li, Zhen-Qiang Yin, Yu-Chun Wu, Xu-Bo Zou, Shuang Wang, Wei Chen,
  Guang-Can Guo, and Zheng-Fu Han.
\newblock Semi-device-independent random-number expansion without entanglement.
\newblock {\em Physical Review A}, 84(3):034301, 2011.

\bibitem{pawlowski2011semi}
Marcin Paw{\l}owski and Nicolas Brunner.
\newblock Semi-device-independent security of one-way quantum key distribution.
\newblock {\em Physical Review A}, 84(1):010302, 2011.

\bibitem{wehner2008lower}
Stephanie Wehner, Matthias Christandl, and Andrew~C Doherty.
\newblock Lower bound on the dimension of a quantum system given measured data.
\newblock {\em Physical Review A}, 78(6):062112, 2008.

\bibitem{li2015detection}
Hong-Wei Li, Zhen-Qiang Yin, Marcin Paw{\l}owski, Guang-Can Guo, and Zheng-Fu
  Han.
\newblock Detection efficiency and noise in a semi-device-independent
  randomness-extraction protocol.
\newblock {\em Physical Review A}, 91(3):032305, 2015.

\bibitem{zhou2016semi}
Yu-Qian Zhou, Fei Gao, Dan-Dan Li, Xin-Hui Li, and Qiao-Yan Wen.
\newblock Semi-device-independent randomness expansion with partially free
  random sources using 3→ 1 quantum random access code.
\newblock {\em Physical Review A}, 94(3):032318, 2016.

\bibitem{mironowicz2014properties}
Piotr Mironowicz, Hong-Wei Li, and Marcin Paw{\l}owski.
\newblock Properties of dimension witnesses and their semidefinite programming
  relaxations.
\newblock {\em Physical Review A}, 90(2):022322, 2014.

\bibitem{levenberg1944method}
Kenneth Levenberg.
\newblock A method for the solution of certain non-linear problems in least
  squares.
\newblock {\em Quarterly of Applied Mathematics}, 2(2):164--168, 1944.

\end{thebibliography}
\end{document}